\documentclass[%
reprint,
showpacs,preprintnumbers,showkeys,
 amsmath,amssymb,
 aps,
pra,
]{revtex4-1}
\usepackage{color}
\usepackage{graphicx}
\usepackage{epstopdf}
\usepackage{dcolumn}
\usepackage{bm}

\begin{document}


\title{Potential energy surfaces in atomic structure: The role of Coulomb correlation in the ground state of helium}

\author{L. D. Salas and J. C. Arce}
 \email{julio.arce@correounivalle.edu.co}
\affiliation{
Departamento de Qu\'imica, Universidad del Valle, A.A. 25360, Cali, Colombia
}%

\date{\today}

\begin{abstract}

For the $S$ states of two-electron atoms, we introduce an exact and unique factorization of the internal eigenfunction in terms of a marginal amplitude, which depends functionally on the electron-nucleus distances $r_1$ and $r_2$, and a conditional amplitude, which depends functionally on the interelectronic distance $r_{12}$ and parametrically on $r_1$ and $r_2$.
Applying the variational principle, we derive pseudoeigenvalue equations for these two amplitudes, which cast the internal Schr\"odinger equation in a form akin to the Born-Oppenheimer separation of nuclear and electronic degrees of freedom in molecules.
The marginal equation involves an effective radial Hamiltonian, which contains a nonadiabatic potential energy surface that takes into account all interparticle correlations in an averaged way, and whose unique eigenvalue is the internal energy.
At each point $(r_1,r_2)$, such surface is, in turn, the unique eigenvalue in the conditional equation.
Employing the ground state of He as prototype, we show that the nonadiabatic potential energy surface affords a molecularlike interpretation of the structure of the atom, and aids in the analysis of energetic and spatial aspects of the Coulomb correlation, in particular correlation-induced symmetry breaking and quantum phase transition.
\end{abstract}

\pacs{31.10.+z, 31.15.V-, 31.50.-x}
\keywords{atomic shape, Coulomb correlation, helium atom, marginal-conditional factorization, nonadiabatic potential energy surface.}

\maketitle

\section{\label{sec:introduction} Introduction}

The notion that molecules and other atomic aggregates possess well-defined shapes, arguably the most basic paradigm of chemistry, is rooted in the classical structural theory \cite{Woolley1978,Claverie1980}. In the quantum-mechanical framework, this notion is \textit{put in} on the basis of the topography of a Born-Oppenheimer (BO) potential energy surface (PES) \cite{Sutcliffe2012}.

In atomic physics, the success of the independent-particle central-field model led, for a long time, to the view that atoms are essentially spherical \cite{Berry1989-95}. However, the discovery that the intrashell supermultiplets in the double-excitation spectra of He can be empirically assigned to collective rotational-like and bendinglike  motions of the electrons \cite{Kellman1980}, analogous to the rovibrational motions of a nonrigid linear $XYX$ molecule, challenged this view \cite{Berry1989-95}. Afterwards, it was found that more complex atoms can have nonspherical shapes even in their ground states \cite{Bao1994}. An analogous situation, referred to as ``crystallization", arises in ``artificial atoms", which are mesoscopic systems constituted by electrons confined in semiconductor quantum dots \cite{Puente2004}.

In nuclear physics, it was early recognized that nuclei can deviate from the spherical shape \cite{Nilsson1995}. Moreover, the concept of a BO PES was introduced, in analogy with the molecular case, albeit in the limited context of the empirical liquid drop model, with the roles of the adiabatic degrees of freedom played by the parameters that specify the surface deformation \cite{Bohr1939,Moller2009}.

Without recourse to the BO approximation, for any collection of particles, elements of structure can still be extracted from the patterns in suitably defined probability densities obtained from all-particle  wavefunctions. For example, such densities reveal that in some states of few-valence-electron atoms \cite{Berry1989-95,Bao1994,Berry1981-87} and other few-body systems \cite{Matyus2011,Becerra2013,Birman2013} the correlations can induce such a collective behavior of the particles that the system acquires a moleculelike shape.
Nevertheless, this probability-based notion of ``quantum structure" is not as far reaching as the BO-PES-based notion of classical structure commonly employed in the molecular sciences \cite{Woolley1978,Claverie1980}.

Hence, it is highly desirable to generalize the concept of PES, starting from the quantum-mechanical Hamiltonian, to any assembly of particles, since this would allow the transfer of  chemical-like notions, e.g., bond length, bond angle, and transition state, together with the conceptual and technical apparatus of quantum chemistry and molecular spectroscopy, e.g., the Franck-Condon principle, to other realms, thereby providing a unified treatment of atoms in molecules \cite{Berry1983,Kellman1997}, electrons in atoms \cite{Berry1983,Kellman1997,Essen1977}, quasiparticles in nanostructures \cite{Puente2004,Birman2013}, nucleons in nuclei \cite{Birman2013,Villars1987}, and even quarks in baryons \cite{Richard1992}.
At first sight, this program might seem to be hampered by the fact that an adiabatic separation of degrees of freedom, the basic tenet of the BO approximation \cite{Sutcliffe2012,Essen1977,Villars1987}, is not always possible. However, in this paper we rigorously prove that such generalization is indeed feasible,
by introducing an exact and unique marginal-conditional factorization (MCF) of the wavefunction \cite{Hunter1975}.

As a prototype, we show how to define a \textit{nonadiabatic} PES (NAPES) for the radial motions of the electrons in $S$ states of two-electron atoms, which amounts to formulating an \emph{exact}  central-field model for these states.
For the case of the ground state of He, the topography of this surface allows us to extract a classical molecularlike interpretation of the structure of the atom. In addition, we analyze the contributions of the different aspects of the electron correlation to this topography. Furthermore, we show that this NAPES provides a convenient conceptual framework for addressing issues of correlation-induced symmetry breaking and quantum phase transition.

\maketitle

\section{\label{sec:formalism} Formalism}

For the $S$ states of two-electron atoms, the internal Schr\"odinger equation,
%
\begin{equation}\label{eq:SE}
\hat{H}\Phi(r_1,r_2,r_{12})=E\Phi(r_1,r_2,r_{12}),
\end{equation}
determines the nonrelativistic energy, $E$ \cite{Hylleraas1929-64}. Considering the center of mass located at the nucleus, the internal coordinates $r_1,r_2,r_{12}$, with $r_1,r_2$ being the electron-nucleus distances and $r_{12}$ being the interelectronic distance, determine the shape and size of the electron-nucleus-electron triangle. Such coordinates are independent, except that they are constrained by the triangle condition $|r_1-r_2|\leq r_{12}\leq r_1+r_2$.
The volume element of this configuration space is $dV=r_1r_2r_{12}dr_1dr_2dr_{12}$. In atomic units, the fixed-nucleus Hamiltonian is given by
\begin{eqnarray}\label{eq:Hamiltonian}
\hat{H}=&-&\sum_{i=1}^2\left(\frac{1}{2}\frac{\partial^2}{\partial r_i^2}+\frac{1}{r_i}\frac{\partial}{\partial r_i}+\frac{Z}{r_i}\right)\nonumber\\
&-&\left(\frac{\partial^2}{\partial r_{12}^2}+\frac{2}{r_{12}}\frac{\partial}{\partial r_{12}}-\frac{1}{r_{12}}\right)\nonumber\\
&-&\sum_{i\neq j}^2\frac{r_i^2-r_j^2+r_{12}^2}{2r_ir_{12}}\frac{\partial^2}{\partial r_i \partial r_{12}}.
\end{eqnarray}
Evidently, in this coordinate system there appear kinetic couplings, i.e., terms containing derivatives with respect to $r_{12}$, besides the original Coulomb potential coupling, $r_{12}^{-1}$.
 
We wish to construct a PES for the radial degrees of freedom, $r_1,r_2$, which we will accomplish by averaging over the $r_{12}$ variable, in the following way. The internal distribution function is given by
\begin{equation}\label{eq:dist}
D(r_1,r_2,r_{12})=r_1r_2r_{12}|\Phi(r_1,r_2,r_{12})|^2.
\end{equation}
According to Bayes' rule, this function can be factorized as
\begin{equation}\label{eq:Bayes}
D(r_1,r_2,r_{12})=D_m(r_1,r_2)D_c(r_{12}|r_1,r_2),
\end{equation}
where
\begin{equation}\label{eq:mardist}
D_m(r_1,r_2):=\int^{r_1+r_2}_{|r_1-r_2|}dr_{12}D(r_1,r_2,r_{12})
\end{equation}
is the marginal distribution function for finding the electrons at $(r_1,r_2)$ irrespective of $r_{12}$, and $D_c(r_{12}|r_1,r_2)$ is the conditional distribution function for finding the electrons at $r_{12}$ provided that they are found at $(r_1,r_2)$. (Here and henceforth, when we speak of the probability of finding the electrons at a point, we actually mean in an infinitesimal neighborhood around that point.) The normalization of $D(r_1,r_2,r_{12})$,
\begin{equation}\label{eq:normdist}
\int_0^{\infty}dr_1\int_0^{\infty}dr_2\int_{|r_1-r_2|}^{r_1+r_2}dr_{12}D(r_1,r_2,r_{12})=1,
\end{equation}
automatically implies the normalization of $D_m(r_1,r_2)$,
\begin{equation}\label{eq:normmar}
\int_0^{\infty}dr_1\int_0^{\infty}dr_2D_m(r_1,r_2)=1,
\end{equation}
and the \emph{local} normalization of $D_c(r_{12}|r_1,r_2)$,
\begin{equation}\label{eq:normcon}
\int_{|r_1-r_2|}^{r_1+r_2}dr_{12}D_c(r_{12}|r_1,r_2)=1.
\end{equation}

Following Hunter \cite{Hunter1975}, we introduce the MCF of the internal eigenfunction
\begin{equation}\label{eq:MCF}
\Phi(r_1,r_2,r_{12})=\psi(r_1,r_2)\chi(r_{12}|r_1,r_2),
\end{equation}
by defining marginal and conditional \emph{amplitudes} \cite{Gross2012-14,Arce2012}
\begin{eqnarray}\label{eq:defpsi}
\psi(r_1,r_2)&:=& e^{i\alpha(r_1,r_2)}\left(\int_{|r_1-r_2|}^{r_1+r_2}dr_{12}r_{12}|\Phi(r_1,r_2,r_{12})|^2\right)^{1/2}\nonumber\\
&\equiv& e^{i\alpha(r_1,r_2)}\left\langle\Phi|\Phi\right\rangle^{1/2},
\end{eqnarray}
\begin{equation}\label{eq:defchi}
\chi(r_{12}|r_1,r_2):=e^{-i\alpha(r_1,r_2)}
\frac{\Phi(r_1,r_2,r_{12})}{\left\langle\Phi|\Phi\right\rangle^{1/2}},
\end{equation}
such that
\begin{equation}\label{eq:qmardist}
D_m(r_1,r_2)=r_1r_2|\psi(r_1,r_2)|^2,
\end{equation}
\begin{equation}\label{eq:qcondist}
D_c(r_{12}|r_1,r_2)=r_{12}|\chi(r_{12}|r_1,r_2)|^2.
\end{equation}
[From Eq. (\ref{eq:defpsi}) onwards, angular brackets express integrals over $r_{12}$ with the Jacobian $r_{12}$.]
The normalization conditions (\ref{eq:normdist})--(\ref{eq:normcon}) now read
\begin{equation}\label{eq:normPhi}
\int_0^{\infty}dr_1r_1\int_0^{\infty}dr_2r_2\left\langle \Phi|\Phi\right\rangle=1,
\end{equation}
\begin{equation}\label{eq:normpsi}
\int_0^{\infty}dr_1r_1\int_0^{\infty}dr_2r_2|\psi|^2=1,
\end{equation}
\begin{equation}\label{eq:normchi}
\left\langle \chi|\chi\right\rangle=1.
\end{equation}

The following observations about the MCF [Eqs. (\ref{eq:MCF})-(\ref{eq:normchi})] are in order. First, it does not presuppose an adiabatic separation of the degrees of freedom, i.e., that $r_1$ and $r_2$ are ``slow" in comparison with $r_{12}$.
Second, even if the last summation in Eq. (\ref{eq:Hamiltonian}), which explicitly couples all the variables, were neglected, $\Phi(r_1,r_2,r_{12})$ still could not be exactly factorized as, say, $\psi(r_1,r_2)\xi(r_{12})$, because of the triangle condition.
Third, the phase $e^{i\alpha(r_1,r_2)}$, with $\alpha(r_1,r_2)$ real, is arbitrary. Thus, given the exchange symmetry of $\Phi$, this phase  can be chosen to set the exchange symmetries of $\psi$ and $\chi$, in the following way: Noting that $\left\langle\Phi|\Phi\right\rangle^{1/2}$ is always symmetric, for singlet states, where $\Phi$ must be symmetric, $\alpha$ can be chosen as symmetric (antisymmetric) to make both $\psi$ and $\chi$ symmetric (antisymmetric); for triplet states, where $\Phi$ must be antisymmetric, $\alpha$ can be chosen as symmetric (antisymmetric) to make $\psi$ symmetric (antisymmetric) and $\chi$ antisymmetric (symmetric).
Fourth, the local normalization of $\chi$ (Eq. (\ref{eq:normchi})) guarantees it to be nontrivial and unique, within the phase  $\alpha(r_1,r_2)$ \cite{Gross2012-14}.
Fifth, $\chi$ is not \emph{globally} normalizable despite the fact that $\Phi$ is, since  $\int_0^{\infty}dr_1r_1\int_0^{\infty}dr_2r_2\left\langle \chi|\chi\right\rangle=\mathcal{V}\rightarrow\infty$, with $\mathcal{V}$ the volume of the $\{r_1,r_2\}$ subspace \cite{Arce2012}.
Finally, if $\psi(r_1,r_2)$ had a node at, say, $r_1=a$ then $\Phi(a,r_2,r_{12})$ would have to vanish at all $r_{12}$ [see Eq. (\ref{eq:defpsi})]. Moreover, for $\chi(r_{12}|r_1,r_2)$ to remain finite as $r_1\rightarrow a$, $\Phi(r_1,r_2,r_{12})$ would have to approach zero faster than $\left\langle \Phi|\Phi\right\rangle^{1/2}$ at all $r_{12}$ in this limit [see Eq. (\ref{eq:defchi})], which is impossible. This constitutes a proof that marginal amplitudes must be nodeless, alternative to the one presented by Hunter \cite{Hunter1981}.

We derive the equations that govern $\psi$ and $\chi$ from the variational principle, as follows. First, we set up the constrained functional \cite{Gross2012-14}
\begin{eqnarray}\label{eq:functional}
F[\Phi]&\equiv&\int_0^{\infty}dr_1r_1\int_0^{\infty}dr_2r_2\left\langle\Phi|\hat{H}|\Phi\right\rangle\nonumber\\
&-&\int_0^{\infty}dr_1r_1\int_0^{\infty}dr_2r_2\lambda(r_1,r_2)\left(\left\langle \chi|\chi\right\rangle-1\right)\nonumber\\
&-&\epsilon\left(\int_0^{\infty}dr_1r_1\int_0^{\infty}dr_2r_2|\psi|^2-1\right),
\end{eqnarray}
where the first term is the expectation value of the energy, the second term ensures the local normalization of $\chi(r_{12}|r_1,r_2)$ at every point of the $\{r_1,r_2\}$ subspace, and the third term ensures the normalization of $\psi(r_1,r_2)$, with $\lambda(r_1,r_2)$ and $\epsilon$ being Lagrange multipliers. Then, we impose the extremization condition $\delta F=0$, which yields, after a convenient rearrangement,
\begin{equation}\label{eq:mareq}
\left(\hat{T}_{0}+U(r_1,r_2)\right)\psi(r_1,r_2)=\epsilon\psi(r_1,r_2),
\end{equation}
\begin{equation}\label{eq:coneq}
\hat{\Omega}_{\psi}(r_1,r_2)\chi(r_{12}|r_1,r_2)
=U(r_1,r_2)\chi(r_{12}|r_1,r_2),
\end{equation}
where
\begin{equation}\label{eq:T0} \hat{T}_0\equiv-\sum_i\left(\frac{1}{2}\frac{\partial^2}{\partial r_i^2}+\frac{1}{r_i}\frac{\partial}{\partial r_i}\right),
\end{equation}
\begin{equation}\label{eq:defOmega}
\hat{\Omega}_{\psi}(r_1,r_2)\equiv\hat{H}
-\sum_{i\neq j}^2\left(\frac{1}{\psi}\frac{\partial\psi}{\partial r_i}\right)
\left(\frac{r_i^2-r_j^2+r_{12}^2}{2r_ir_{12}}\frac{\partial}{\partial r_{12}}+\frac{\partial}{\partial r_i}\right).
\end{equation}
Here, the notation $\hat{\Omega}_{\psi}(r_1,r_2)$ indicates that this operator contains $\psi(r_1,r_2)$ and acts also on the $r_1,r_2$ variables. The fact that in Eq. (\ref{eq:coneq}) $r_1$ and $r_2$ play the roles of parameters can be made more explicit by rewriting its left-hand side as $\hat{\Omega}_{\psi}(r_1',r_2')\chi(r_{12}|r_1',r_2')\vline_{r_1'=r_1,r_2'=r_2}$, which means that $r_1$ and $r_2$ are fixed only after the operator has acted on $\chi$.

Equations (\ref{eq:mareq}) and (\ref{eq:coneq}) constitute a pair of \emph{exact} coupled pseudoeigenvalue equations with the following characteristics. First, since the operators on their left-hand sides are Hermitian the eigenvalues $\epsilon$  and $U(r_1,r_2)$ are real.
Second, since $\psi$ and $\chi$ are unique up to a phase for a given $\Phi$, each one of them possesses only one acceptable solution.
Third, the presence of the logarithmic derivatives of $\psi$ in the second term at the right-hand side of Eq. (\ref{eq:defOmega}) makes them nonlinear, which implies that their solution requires an iterative self-consistent scheme. (For mathematical caveats about this kind of problem, see Ref. \cite{Jecko2015}.)
Finally, the eigenvalue $U(r_1,r_2)$ and the Lagrange multiplier $\lambda(r_1,r_2)$ are related by
\begin{equation}\label{eq:defU}
U(r_1,r_2)=
\frac{\lambda(r_1,r_2)}{\rho_m(r_1,r_2)}-\frac{\hat{T}_{0}\psi(r_1,r_2)}{\psi(r_1,r_2)}.
\end{equation}
Consequently, by substituting Eq. (\ref{eq:defU}) into Eq. (\ref{eq:mareq}) and taking into account Eq. (\ref{eq:normpsi})
we see that
$\epsilon=\int_0^{\infty}dr_1r_1\int_0^{\infty}dr_2r_2\lambda(r_1,r_2)$. Furthermore, with some additional manipulation we obtain that $\epsilon=\int_0^{\infty}dr_1r_1\int_0^{\infty}dr_2r_2\left\langle \Phi|\hat{H}|\Phi\right\rangle=E$. Therefore, $\lambda(r_1,r_2)=\left\langle \Phi|\hat{H}|\Phi\right\rangle$ can be interpreted as a local energy, i.e., the energy of the system when the electrons are positioned at $(r_1,r_2)$.
  
In Eq. (\ref{eq:mareq}) $\hat{T}_0$, as given by Eq. (\ref{eq:T0}), represents the kinetic energy of two electrons in a central field. On the other hand, according to Eq. (\ref{eq:coneq}) $U(r_1,r_2)=\left\langle\chi|\hat{\Omega}_{\psi}(r_1,r_2)|\chi\right\rangle$, so this function carries all the information about the electron-nucleus attractions, electron-electron repulsion, and kinetic couplings, averaged over $r_{12}$. Hence, in Eq. (\ref{eq:mareq}) $U(r_1,r_2)$ plays the role of an effective radial potential that, nonetheless, correlates the electrons fully. Consequently, Eq. (\ref{eq:mareq}) constitutes an \emph{exact} central-field model for $S$ states.
In addition, since this is a Schr\"odinger equation for $\sqrt{D_m(r_1,r_2)/r_1r_2}$ (see Eq. (\ref{eq:qmardist})), we observe that the energy is a functional of the marginal distribution function, which from now on we will call the radial distribution.

Equations (\ref{eq:mareq}) and (\ref{eq:coneq}) are arranged in a form analogous to Hunter's nonadiabatic electronic-nuclear separation \cite{Hunter1975,Gross2012-14,Arce2012,Cederbaum2013-2014}, which, in turn, is a sort of exact version of the BO approximation. Thus, $r_{12}$ and $r_1,r_2$ are the analogs of the electronic and the nuclear coordinates, respectively, and $\hat{\Omega}$,  $\hat{T}_0$ and $U$ are the analogs of the clamped-nuclei Hamiltonian (with the electron-nucleus attractions included), the nuclear kinetic energy operator, and the molecular NAPES. Consequently, we will refer to our $U(r_1,r_2)$ as the \emph{atomic} NAPES. By the same token, $\chi$ and $\psi$ are analogous to the electronic and the vibrational eigenfunctions.
Nevertheless, it must be kept in mind that, due to the uniqueness of the MCF, there can be only one marginal (``vibrational") eigenfunction, with energy $E$, associated with this atomic NAPES, in contrast with the molecular BO case, where the PES can sustain several vibrational states with different energies.

\maketitle

\section{\label{sec:calculations} Illustrative Calculations and Discussion}

We applied the foregoing formalism to the ground state of He. The solution to the nonlinear system of Eqs. (\ref{eq:mareq}) and (\ref{eq:coneq}) is a difficult problem \cite{Jecko2015} that lies beyond the scope of this work. Since our main goal is to learn about the topography of the atomic NAPES and its dependence upon the interparticle correlations, we employed these equations for the purposes of analysis only. Therefore, we extracted approximate marginal and conditional amplitudes, using Eqs. (\ref{eq:defpsi}) and (\ref{eq:defchi}), choosing $\alpha\equiv 0$, from variationally optimized trial functions, and then evaluated approximate atomic NAPES's by   $U\approx\left\langle\chi|\hat{\Omega}_{\psi}|\chi\right\rangle$, in accordance with Eq. (\ref{eq:coneq}). 

In order of increasing account of the Coulomb correlation, the optimized trial functions selected were
\begin{equation}\label{eq:Ke}
\Phi(r_{1},r_{2},r_{12})=Ne^{-\zeta(r_{1}+r_{2})},
\end{equation}
with $N=13.59$, $\zeta=27/16$;
\begin{equation}\label{eq:Li}
\Phi(r_{1},r_{2},r_{12})=Ne^{-\zeta(r_{1}+r_{2})}(1+c_1r_{12}),
\end{equation}
with $N=12.26$, $\zeta=1.857270$, $c_1=0.390807$; and
\begin{eqnarray}\label{eq:Hy}
\Phi(r_{1},r_{2},r_{12})= &N&e^{-\zeta(r_{1}+r_{2})}[1+c_1r_{12}+c_2r_{12}^2
+c_3(r_{1}+r_{2})\nonumber\\
&+&c_4(r_{1}+r_{2})^2+c_5(r_{1}-r_{2})^2],
\end{eqnarray}
with $N=12.27$, $\zeta=1.755656$, $c_1=0.337294$, $c_2=-0.037024$, $c_3=-0.145874$, $c_4=0.023634$, $c_5=0.112519$.
Expression (\ref{eq:Ke}) is the familiar Kellner uncorrelated function, whereas expressions  (\ref{eq:Li}) and (\ref{eq:Hy}) are of the Hylleraas type, with linear and quadratic polynomial correlation factors, respectively \cite{Hylleraas1929-64}.
For these trial functions the expectation values of the energy turn out to be $\langle E\rangle=-2.848,-2.891,-2.903$ hartree, respectively  \cite{Hylleraas1929-64}, which increasingly approach the reported ``exact"  energy \cite{Nakashima2007}.
Of course, much better trial functions than (\ref{eq:Hy}) can be devised \cite{Nakashima2007}, but this level of approximation suffices for our largely qualitative analysis.

Before presenting the results, we must qualitatively compare the spatial behavior of the functions (\ref{eq:Ke})-(\ref{eq:Hy}).
First, we recall that, strictly, the correlation factor in a trial function should go to a constant as $r_{12}\rightarrow\infty$ (which can occur only when $r_1\rightarrow\infty$ and/or $r_2\rightarrow\infty$),  because in this limit the electrons are uncorrelated. Hence, the only  function that fulfills this condition is (\ref{eq:Ke}). In fact, the correlation factors of functions (\ref{eq:Li}) and  (\ref{eq:Hy}) diverge in this limit. (However, the full trial functions remain well behaved because the exponential factor damps the divergence.) Thus, in regions of large $r_1$ and/or large $r_2$ the uncorrelated  function (\ref{eq:Ke}) actually provides a much better approximation to the exact wavefunction. On the other hand, close to the nucleus the correlated functions (\ref{eq:Li}) and (\ref{eq:Hy}) provide better approximations. Very close to the nucleus, function (\ref{eq:Hy}) should provide a better approximation than function (\ref{eq:Li}), but at sufficiently long distance from the nucleus the situation should reverse, since the divergence of (\ref{eq:Hy}) is stronger. Hence, we will confine our comparison of the NAPES's extracted from these functions to a region relatively close to the nucleus, which is where the electron correlation plays an important role, anyway.
Second, the correlation factor of function (\ref{eq:Hy}) also has corrections in $r_1$ and $r_2$, which (\ref{eq:Li}) does not have.
Therefore, due to the interplay of these characteristics, at a particular point $(r_1,r_2)$ it is difficult to predict which NAPES will be more accurate, although, evidently, the one evaluated with function (\ref{eq:Hy}) should be more accurate overall, since it provides the best energy.

Trial functions with polynomial correlation factors, in spite of their wrong asymptotic behaviors, are advantageous to us because they permit analytical evaluations of $\psi$, $\chi$ and $U$. For this task, we employed \textit{Mathematica 10.3.1} \cite{Mathematica}. However, we do not provide the expressions here, since some of them are too formidable. As a check, we also reevaluated the energies by means of the expression 
$\langle E\rangle=\int_0^{\infty}dr_1r_1\int_0^{\infty}dr_2r_2\psi^*(r_1,r_2)
\left(\hat{T}_0+U(r_1,r_2)\right)\psi(r_1,r_2)$ (see Eq. (\ref{eq:mareq})), which had to be performed numerically, employing the same software; the values obtained turned out to be the same as the ones reported above.

\begin{figure}
\includegraphics[scale=0.45]{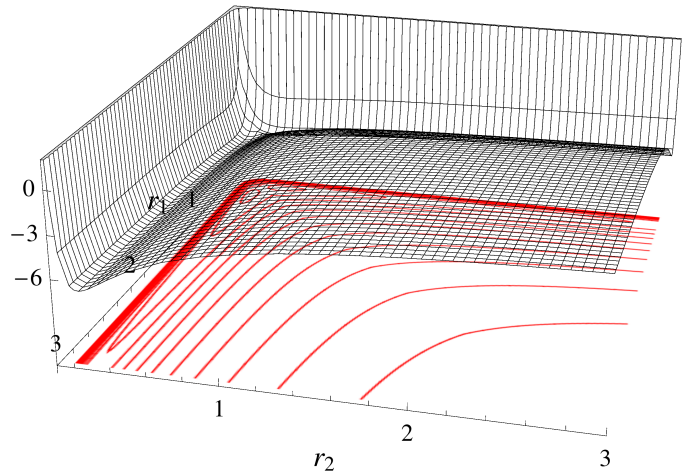}
\caption{\label{fig:Fig-1_Salas} The atomic NAPES, $U(r_1,r_2)$, and its contour map, obtained from the trial function (\ref{eq:Hy}). All quantities are given in atomic units. The innermost contour has a value of $-7.5$ hartree and contour values increase outwards in intervals of 0.5.}
\end{figure}

Figure \ref{fig:Fig-1_Salas} displays the approximate atomic NAPES evaluated with the trial function (\ref{eq:Hy}). Several features stand out. First, $U$ has negative values everywhere, except for the steep repulsive walls at small $r_1$ or $r_2$. Second, it contains a ridge along $r_1=r_2$, which asymptotically goes to zero very slowly. (Strictly speaking, exactly at $r_1=r_2$ the value of $U$ is undetermined, an artifact of the approximate $\Phi$ employed.
We obtained these values by a simple interpolation. The same observation applies to the NAPES that will be obtained later from the trial function (\ref{eq:Li})).
Third, this ridge separates two basins with minima of $U_e=-7.73$ hartree positioned at $r_{e,i}=0.26,r_{e,j}=0.43$ bohr, and asymptotic values of $-5.57$ hartree. Fourth, there is a saddle point of $U^\ddagger=-7.42$ hartree positioned at $r_1^\ddagger=r_2^\ddagger=0.32$ bohr, which, along the minimum-energy path, lies at the top of a barrier of height $U_b\equiv U^\ddagger-U_e=0.31$ hartree located in between the two minima.

A classical molecularlike interpretation of this topography is as follows. The atomic NAPES is associative, as it should since we are dealing with a bound state. The minima correspond to two versions of the same equilibrium structure, with the two electrons on respective ``Bohrian" orbits of radii 0.26 and 0.43 bohr. These versions are interconvertible by a degenerate rearrangement through the saddle point \cite{Bunker1998}, the latter corresponding to a transition structure with the two electrons on the same (unstable) orbit of radius 0.32 bohr.
The stable orbits can perform highly anharmonic ``breathing" motions, which are analogous to molecular bond stretching modes. Since the spatial part of the eigenfunction is symmetric under permutation of $r_1$ and $r_2$ (see Eq. (\ref{eq:Hy})), these breathings occur in phase, analogously to a molecular symmetric stretching mode.

\begin{figure}
\includegraphics[scale=0.55]{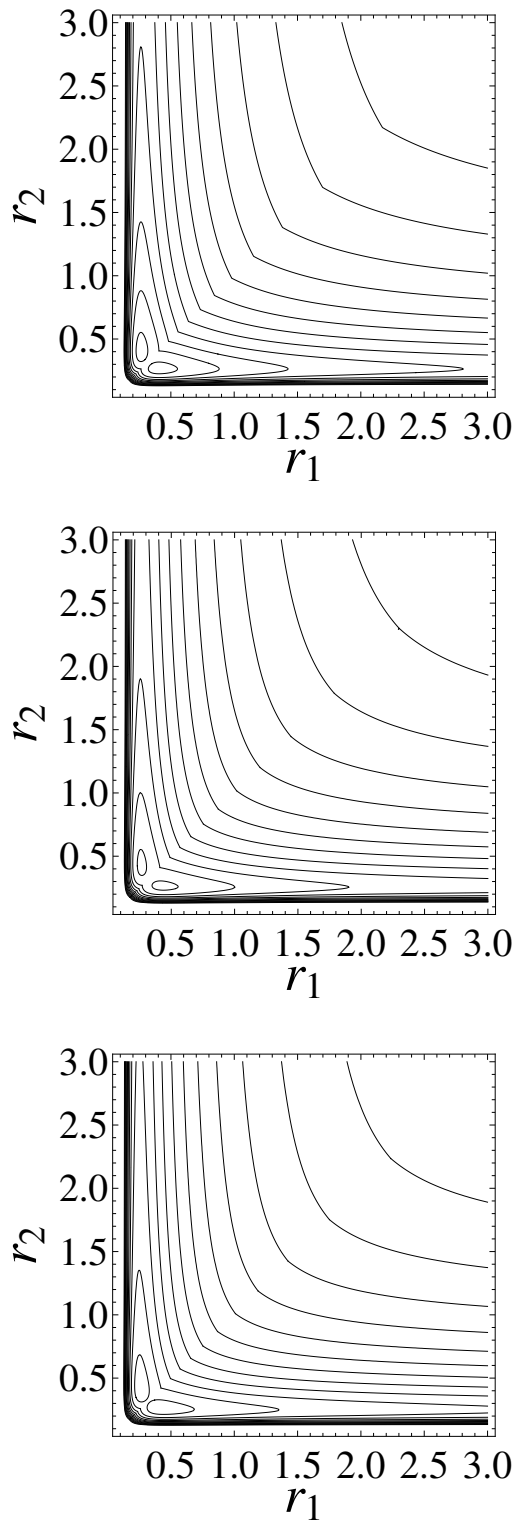}
\caption{\label{fig:Fig-2_Salas} Contour maps of the atomic NAPES's obtained with trial functions carrying different degrees of electron correlation. All quantities are given in atomic units. Top: constant correlation factor (Eq.(\ref{eq:Ke})). Middle: linear correlation factor (Eq.(\ref{eq:Li})). Bottom: quadratic correlation factor (Eq.(\ref{eq:Hy})). The innermost countour has a value of $-7.5$ hartree and contour values increase outwards in intervals of 0.5.}
\end{figure}

To examine the sensitivity of the topography of the NAPES to the degree of accuracy with which the electron correlation is taken into account, we also evaluated NAPES's with the two other trial functions. Figure \ref{fig:Fig-2_Salas} shows the results. (We chose the size of the region exhibited in accordance with the effective range of the radial distribution, as discussed below.) We see that the three surfaces are very similar, which means that the topography of the NAPES is not very sensitive to the degree of correlation in the wavefunction. As the electron correlation in the trial function increases, the only trend we can discern is that the basins become more anharmonic, which implies that the amplitudes of the breathings of the orbits get larger.

\begin{figure}
\includegraphics[scale=0.55]{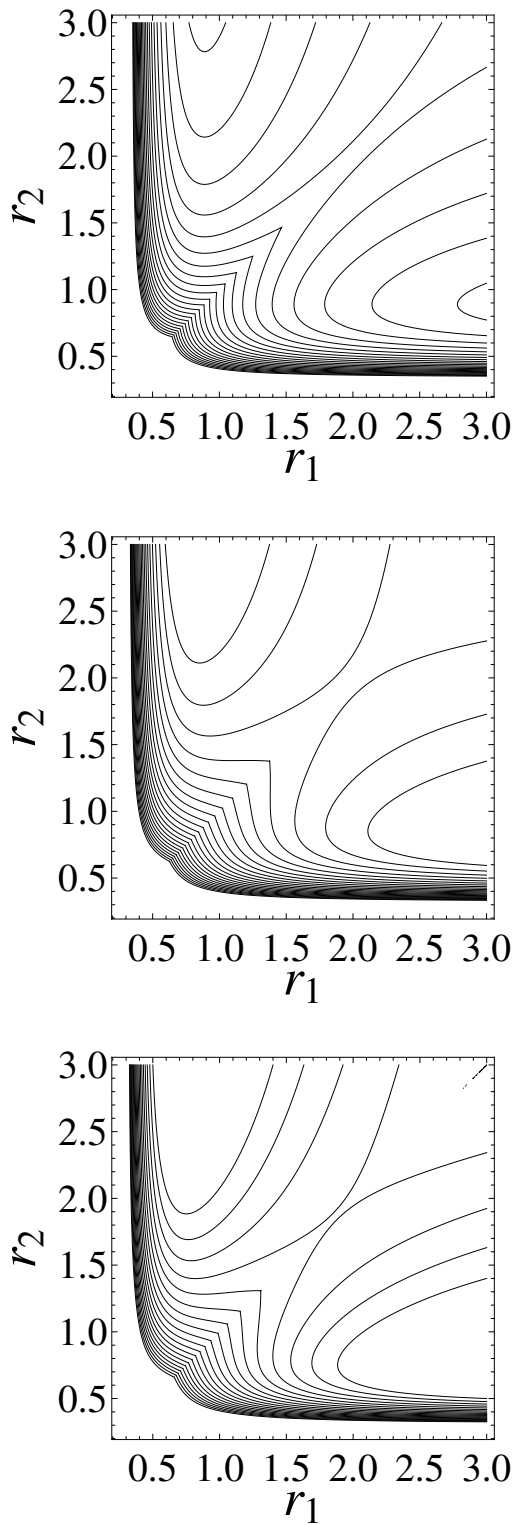}
\caption{\label{fig:Fig-3_Salas} Contour maps of the atomic NAPES's  obtained after removing the electron-nuclear attractions from $\hat{\Omega}$, for the same cases of Fig. \ref{fig:Fig-2_Salas}. All quantities are given in atomic units. The innermost countour has a value of $-0.37$ hartree and contour values increase outwards in intervals of 0.05.}
\end{figure}

\begin{figure}
\includegraphics[scale=0.55]{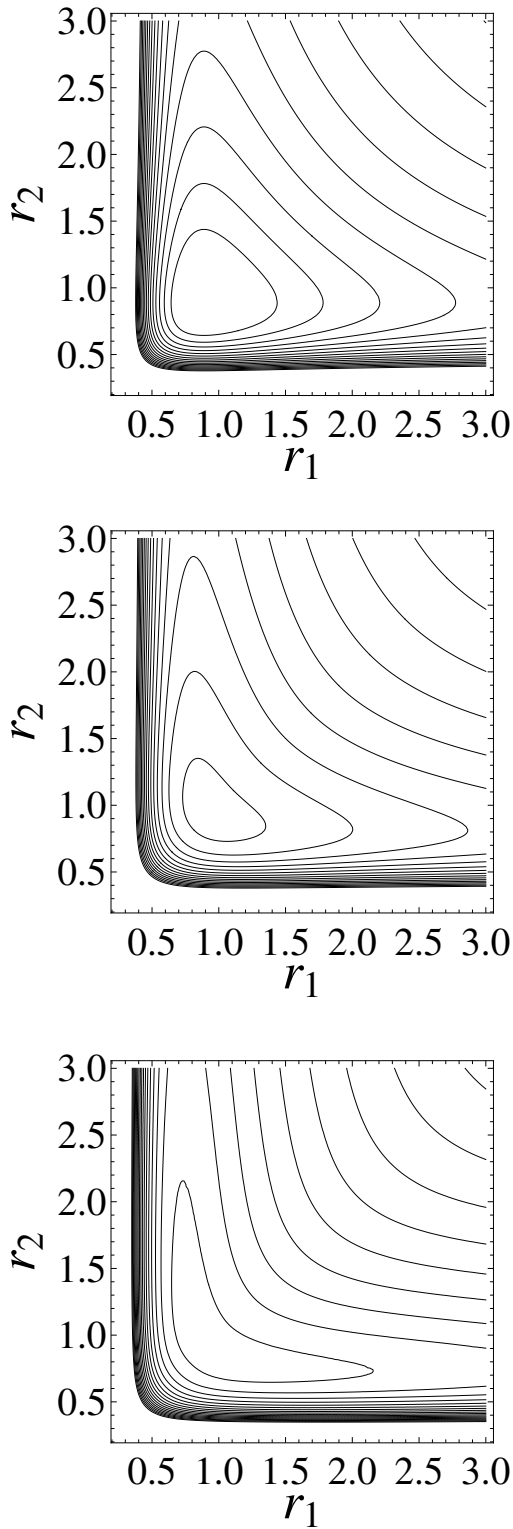}
\caption{\label{fig:Fig-4_Salas} Contour maps of the atomic NAPES's obtained after removing the electron-nuclear attractions and electron-electron repulsion from $\hat{\Omega}$, for the same cases of Fig. \ref{fig:Fig-2_Salas}. All quantities are given in atomic units. The innermost countour has a value of $-0.88$ hartree and contour values increase outwards in intervals of 0.05.}
\end{figure}

To assess the roles of the Coulomb interactions in shaping the topography of the atomic NAPES, first we calculated surfaces with the electron-nucleus  attractions, $-2r_i^{-1}$, removed from $\hat{\Omega}$ [see Eqs. (\ref{eq:defOmega}) and (\ref{eq:Hamiltonian})], for the three trial functions. As observed in Fig. \ref{fig:Fig-3_Salas}, the wells disappear, i.e., the basins become valleys, and, consequently, $U$ becomes dissociative. Moreover, as the degree of correlation carried by the trial function increases, the maximum on the ridge slightly shifts towards shorter distance from the nucleus.
Then, we calculated surfaces with both $-2r_i^{-1}$ and $r_{12}^{-1}$ removed from $\hat{\Omega}$, which means that the only couplings remaining are the kinetic ones. Now, in Fig. \ref{fig:Fig-4_Salas} we see that the ridge disappears and a single shallow well remains, whose bottom corresponds to an equilibrium configuration with both electrons on the same orbit of radius $\sim\ $ 0.9 bohr. Furthermore, as the degree of correlation increases, such well becomes more anharmonic. Thus,  the kinetic couplings play a small stabilizing role in the atom, since this surface is associative, and contribute to the anharmonicity of the basins present in the NAPES (see Fig. \ref{fig:Fig-2_Salas}). Naturally, the electron-nucleus attractions play the major stabilizing role, manifested by the presence of the deep wells in the NAPES (see Fig. \ref{fig:Fig-2_Salas}), and the electron-electron repulsion plays a destabilizing role, manifested by the presence of the ridge (see Figs. \ref{fig:Fig-2_Salas} and \ref{fig:Fig-3_Salas}).
Therefore, the Coulomb repulsion is responsible for inducing the symmetry breaking \cite{Birman2013} that gives rise to the split-orbit equilibrium configuration.
Interestingly, the concept of a correlation-induced orbit splitting has been introduced in an \emph{ad hoc} manner by defining inner and outer radial probability densities \cite{Koga2006}.

\begin{figure}
\includegraphics[scale=0.5]{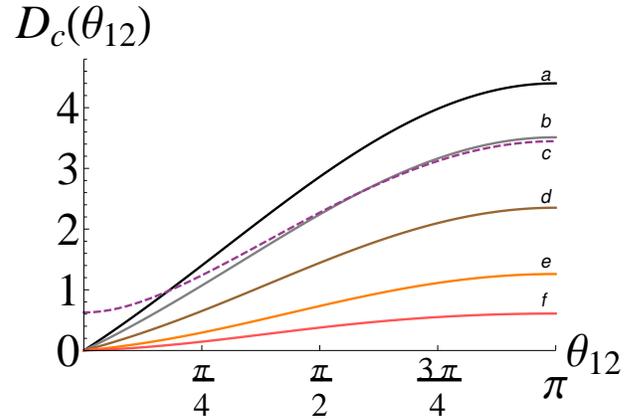}
\caption{\label{fig:Fig-5_Salas} The conditional distribution function, $D_c(\theta_{12}|r_1,r_2)$ (in atomic units), evaluated at (a) $r_1=r_2=0.25$ bohr, (b) $r_1=r_2=0.32$ bohr (the saddle point), (c) $r_i=0.26,r_j=0.43$ bohr (the minimum of either well), (d) $r_1=r_2=0.50$ bohr,  (e) $r_1=r_2=1.00$ bohr, (f) $r_1=r_2=2.00$ bohr.}
\end{figure}

Because of the averaging over $r_{12}$, this atomic NAPES provides explicit information about the radial correlations only. To obtain explicit information about the angular correlations, we turn to the conditional distribution function, $D_c(r_{12}|r_1,r_2)$, which, for convenience of interpretation, we transformed into $D_c(\theta_{12}|r_1,r_2)$ using the relation $r_{12}=(r_1^2+r_2^2-2r_1r_2\cos\theta_{12})^{1/2}$. Figure \ref{fig:Fig-5_Salas} displays this function, extracted from the trial function (\ref{eq:Hy}), evaluated at selected points $(r_1,r_2)$ of the NAPES.
Its behavior reveals the presence of the Coulomb hole: First, the probabilities of finding the system in the collinear electron-nucleus-electron ($\theta_{12}=\pi$) and nucleus-electron-electron ($\theta_{12}=0$) configurations are maximal and minimal, respectively. [Note that the probability of finding the electrons at zero separation vanishes because of the Jacobian $r_{12}$ present in Eq. (\ref{eq:qcondist}).]
Second, the values and curvature of this function decrease with the distance from the nucleus, which must be due to the weakening of the electron-electron correlation as the size of the electron-nucleus-electron triangle grows.

\begin{figure}
\includegraphics[scale=0.28]{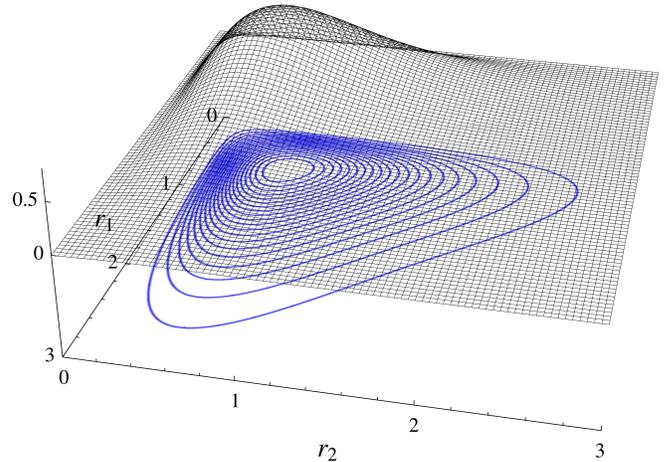}
\caption{\label{fig:Fig-6_Salas} The radial distribution, $D_m(r_1,r_2)$, and its contour map. All quantities are given in atomic units. The innermost countour has a value of 0.66 hartree and contour values decrease outwards in intervals of 0.033.}
\end{figure}

The corresponding radial distribution, $D_m(r_1,r_2)$, is shown in Fig. \ref{fig:Fig-6_Salas}. It turns out to be unimodal, with its maximum located on the ridge, at $r_1=r_2=0.60$ bohr. (By $r_i\sim 3\ $ bohr, this function has practically decayed to zero. This is why above we showed all the surfaces within the region $0\leq r_i\leq 3.0\ $ bohr.)
Hence, quantum mechanics has frustrated the symmetry breaking ``latent" in the NAPES, and placed the most probable configuration with both electrons on the same orbit of radius 0.60 bohr.
Taking into account that the profile of the NAPES along the minimum-energy path looks like a double well, we suspect this has happened because the ``zero-point energy", $E_0\equiv\langle E\rangle -U_e=4.83$ hartree, is very much above the barrier, $U_b=0.31$ hartree. Consequently, the atom behaves analogously to a fluxional molecule. Had $E_0$ been below the barrier, $D_m$ would have displayed two humps, each associated mainly with the breathing of one orbit. It appears that something like this occurs in some of the doubly-excited states of He, for instance the nominal $2s3s\ $ $^1S^e$ state [see Fig. 2 in Ref. \cite{Berry1981-87}(d)], where the electron-electron correlation is more effective and, consequently, the ridge must be higher.

\begin{figure}
\includegraphics[scale=0.55]{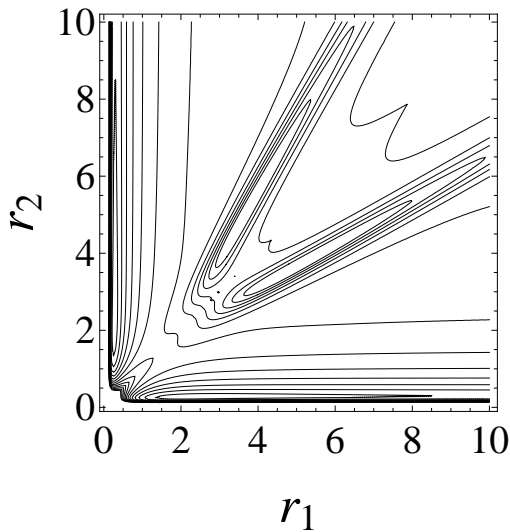}
\caption{\label{fig:Fig-7_Salas} Contour map of the atomic NAPES for a fictitious He atom with the electron-electron repulsion quadrupled. All quantities are given in atomic units. The innermost contour has a value of -4.2 hartree and contour values increase outwards in intervals of 0.55.}
\end{figure}

\begin{figure}
\includegraphics[scale=0.28]{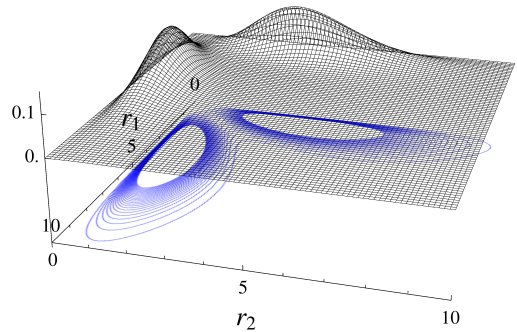}
\caption{\label{fig:Fig-8_Salas} The broken-symmetry radial distribution, $D_m(r_1,r_2)$, and its contour map, for a fictitious He atom with the electron-electron repulsion quadrupled. All quantities are given in atomic units. The innermost countour has a value of 0.0325 hartree and contour values decrease outwards in intervals of 0.0013.}
\end{figure}

To support this suspicion, we considered fictitious atoms with the electron-electron repulsion increased, i.e., with $r_{12}^{-1}$ replaced by $\kappa r_{12}^{-1}$ ($\kappa>1$) in the Hamiltonian (\ref{eq:Hamiltonian}).
For each value of $\kappa$ we used a trial function of the form (\ref{eq:Hy}), with all the parameters reoptimized.
In the NAPES, we observed that, as $\kappa$ increases, hills continuously develop on each side of the ridge and the basins become more anharmonic. Concomitantly, at $\kappa\sim 2.5$ the radial distribution begins to develop two humps along the basins.
Figures \ref{fig:Fig-7_Salas} and \ref{fig:Fig-8_Salas} display the NAPES and $D_m$, respectively, for $\kappa=4$. In this case the energy turns out to be $\langle E\rangle=-1.302$ hartree, which indicates that this fictitious atom is much less stable than the real one, as expected. The minima of the basins are now located at $r_{e,i}=0.27, r_{e,j}=2.28$ bohr with potential values of $U_e=-4.53$ hartree, and the saddle point is now positioned at $r_1^\ddagger=r_2^\ddagger=1.60$ bohr with a potential value of $U^\ddagger=-0.97$ hartree. The barrier height becomes $U_b=3.56$ hartree, which is much higher than the one for the $\kappa=1$ case.
The two maxima of $D_m$ are positioned at $r_i=0.65, r_j=3.64$, very displaced from the potential minima due to the high anharmonicity of the basins. Along the ridge this function practically vanishes. Thus, the most probable configurations now have the electrons on respective orbits of radii 0.65 and 3.64 bohr, with the probability of interconverting by tunneling practically vanishing. These orbits can be envisioned undergoing essentially independent in-phase breathing motions.
Hence, the symmetry breaking latent in the NAPES has become actual, due to the stronger electron-electron correlation \cite{Birman2013}. With confidence, we can associate this phenomenon with the fact that the zero-point energy, $E_0=3.23$ hartree, is now 0.33 hartree below the barrier.



The smooth evolution of the radial distribution from unimodal to bimodal as the Coulomb strength parameter, $\kappa$, increases is the hallmark of a continuous quantum phase transition, where $\kappa$ and the double-well profile of the minimum-energy path are analogous to the temperature and the free energy in thermodynamics, respectively. Phase transitions of this type in Coulomb three-body systems have been characterized \cite{Kais2000}; the atomlike (unimodal) to moleculelike (bimodal) evolution of the particle density in the sequence H$^-\rightarrow$He$\rightarrow$Ps$^-\rightarrow$H$_2^+$ \cite{Matyus2011,Becerra2013} is a case in point.

\maketitle

\section{\label{sec:conclusions} Conclusions and Outlook}

The particular MCF (\ref{eq:MCF}) defined in this work produced an exact central-field model of the two-electron atom, embodied in the atomic NAPES $U(r_1,r_2)$ [see Eq. (\ref{eq:mareq})].
For the ground state of He, we found that the topography of this surface (see Figs. \ref{fig:Fig-1_Salas} and \ref{fig:Fig-7_Salas}), together with the radial distribution (see Figs. \ref{fig:Fig-6_Salas} and \ref{fig:Fig-8_Salas}), provide a convenient framework for discussing issues of electron correlation, symmetry breaking, and quantum phase transition.

Three alternative MCF's can be defined, which allow one to focus on complementary aspects.
First, the MCF $\Phi(r_1,r_2,r_{12})=\psi(r_2,r_{12})\chi(r_1|r_2,r_{12})$ yields the NAPES $U(r_2,\theta_{12})$, which is useful for analyzing the spatial behavior of the Coulomb hole \cite{Boyd1982}. In addition, such NAPES, together with the marginal distribution $D_m(r_2,\theta_{12})$, should also provide a more complete picture of the correlation-induced symmetry breaking underlying the quantum phase transitions that have been observed in Coulomb three-body systems \cite{Matyus2011,Becerra2013,Birman2013,Kais2000}. 
Second, from the MCF $\Phi(r_1,r_2,r_{12})=\psi(r_1)\chi(r_2,r_{12}|r_1)$
the density $\rho(r_2,\theta_{12}|r_1)$ used extensively in an \textit{ad hoc} way by Berry and coworkers \cite{Berry1989-95,Berry1981-87,Berry1983} can be obtained simply as $|\chi(r_2,r_{12}|r_1)|^2$ followed by the transformation $r_{12}\rightarrow\theta_{12}$. In this case, an atomic nonadiabatic potential energy \textit{curve} (NAPEC) $U(r_1)$ can also be extracted \cite{Hunter1986}.
Finally, the MCF  $\Phi(r_1,r_2,r_{12})=\psi(r_{12})\chi(r_1,r_2|r_{12})$ generates the NAPEC $U(r_{12})$, which can be used to generalize the molecular-orbital-like method of Feagin and Briggs, initially devised for the classification of doubly excited states of He \cite{Feagin1986-88}, to cases where the $r_{12}$ coordinate cannot be treated as adiabatic, as in the ground and singly excited states of this atom and other three-particle systems \cite{Hunter1967}.

Further insight can be gained by working with the hyperspherical coordinates $R\equiv (r_1^2+r_2^2)^{1/2}$, $\alpha\equiv\tan^{-1}(r_2/r_1)$ \cite{Lin1995}, instead of the Hylleraas coordinates $r_1,r_2$. In particular, the MCF $\Phi(R,\alpha,\theta_{12})=\psi(R)\chi(\alpha,\theta_{12}|R)$ produces the NAPEC $U(R)$, whose adiabatic counterpart has been extensively used for the classification of states of Coulomb three-particle systems \cite{Lin1995}. (We would like to clarify that what is often called the ``potential surface" in this connection is not a PES in the sense used in this paper, but rather the sum of all the Coulomb potentials appearing in the Hamiltonian expressed in hyperspherical or Jacobi coordinates.)

Lude\~na and coworkers \cite{Becerra2013} have discovered that the topologies of non-BO one-particle nuclear and electron densities for few-body systems depend upon the reference points selected for the definitions of these quantities. This non-uniqueness weakens the notion that elements of molecular structure can be extracted from the patterns exhibited by the densities \cite{Matyus2011}. Here, the MCF can come to the rescue since one can define a density for any particle by  marginalizing the variables associated with the other particles, and, as we have seen, marginal distributions are uniquely defined. All these constitute topics for future research.

We must emphasize that the implementation of this methodology does not rely on finding solutions of coupled marginal and conditional equations, such as (\ref{eq:mareq}) and (\ref{eq:coneq}), which appears to be a very difficult nonlinear problem \cite{Jecko2015}. Instead, these equations are used for extracting and analyzing information contained in wavefunctions that can be generated by other, more practical means, such as the variational trial functions (\ref{eq:Ke})--(\ref{eq:Hy}).
Thus, the extension of this methodology to more complex systems seems quite feasible. One of the issues we plan to address in the near future is whether atoms with more than two valence electrons in their ground states possess well-defined geometrical shapes or behave analogously to fluxional molecules \cite{Bao1994}.


In conclusion, we believe we have taken a first step in developing a research program aimed at unifying the treatment of atoms, molecules, and other collections of quantum particles, where the concept of PES, with the associated notion of geometrical shape, in the sense employed in the molecular sciences, can play a key role.

\begin{acknowledgments}
L.D. Salas acknowledges financial support from Colciencias through a doctoral fellowship.
\end{acknowledgments}
\bigskip


\begin{thebibliography}{22}
\expandafter\ifx\csname natexlab\endcsname\relax\def\natexlab#1{#1}\fi
\expandafter\ifx\csname bibnamefont\endcsname\relax
  \def\bibnamefont#1{#1}\fi
\expandafter\ifx\csname bibfnamefont\endcsname\relax
  \def\bibfnamefont#1{#1}\fi
\expandafter\ifx\csname citenamefont\endcsname\relax
  \def\citenamefont#1{#1}\fi
\expandafter\ifx\csname url\endcsname\relax
  \def\url#1{\texttt{#1}}\fi
\expandafter\ifx\csname urlprefix\endcsname\relax\def\urlprefix{URL }\fi
\providecommand{\bibinfo}[2]{#2}
\providecommand{\eprint}[2][]{\url{#2}}

\bibitem{Woolley1978}
\bibinfo{author}{R.G. Woolley},
\bibinfo{journal}{J. Am. Chem. Soc.}
\textbf{\bibinfo{volume}{100}},
\bibinfo{pages}{1073}
(\bibinfo{year}{1978}).

\bibitem{Claverie1980}
\bibinfo{author}{P. Claverie and S. Diner},
\bibinfo{journal}{Isr. J. Chem.}
\textbf{\bibinfo{volume}{19}},
\bibinfo{pages}{54}
(\bibinfo{year}{1980}).

\bibitem{Sutcliffe2012}
\bibinfo{author}{B.T. Sutcliffe and R.G. Woolley},
\bibinfo{journal}{J. Chem. Phys.}
\textbf{\bibinfo{volume}{137}},
\bibinfo{pages}{22A544-1}
(\bibinfo{year}{2012}).

\bibitem{Berry1989-95}
\bibinfo{author}{R.S. Berry},
\bibinfo{journal}{Contemp. Phys.}
\textbf{\bibinfo{volume}{30}},
\bibinfo{pages}{1}
(\bibinfo{year}{1989});
\bibinfo{author}{in}
\emph{\bibinfo{title}{Structure and Dynamics of Atoms and Molecules: Conceptual Trends}},
edited by \bibinfo{author}{J.L. Calais and E. S. Kryachko}
(\bibinfo{publisher}{Springer Science + Business Media, Dordrecht},
\bibinfo{year}{1995}) p 155.

\bibitem{Kellman1980}
\bibinfo{author}{M.E. Kellman and D.R. Herrick},
\bibinfo{journal}{Phys. Rev. A}
\textbf{\bibinfo{volume}{22}},
\bibinfo{pages}{1536}
(\bibinfo{year}{1980}).

\bibitem{Bao1994}
\bibinfo{author}{C.G. Bao},
\bibinfo{journal}{J. Phys. B: At. Mol. Opt. Phys.}
\textbf{\bibinfo{volume}{26}},
\bibinfo{pages}{4671}
(\bibinfo{year}{1993});
\bibinfo{journal}{Phys. Rev. A}
\textbf{\bibinfo{volume}{50}},
\bibinfo{pages}{2182}
(\bibinfo{year}{1994}).

\bibitem{Puente2004}
\bibinfo{author}{A. Puente, L. Serra, and R.G. Nazmitdinov},
\bibinfo{journal}{Phys. Rev. B}
\textbf{\bibinfo{volume}{69}},
\bibinfo{pages}{125315}
(\bibinfo{year}{2004}).

\bibitem{Nilsson1995}
\bibinfo{author}{S.G. Nilsson and I. Ragnarsson},
\bibinfo{title}{\emph{Shapes and Shells in Nuclear Structure}}
(\bibinfo{publisher}{Cambridge University Press, Cambridge},
\bibinfo{year}{1995}).

\bibitem{Bohr1939}
\bibinfo{author}{N. Bohr and J.A. Wheeler},
\bibinfo{journal}{Phys. Rev.}
\textbf{\bibinfo{volume}{56}},
\bibinfo{pages}{426}
(\bibinfo{year}{1939}).

\bibitem{Moller2009}
\bibinfo{author}{P. M\"oller, A.J. Sierk, T. Ichikawa, A. Iwamoto, R. Bengtsson, H. Uhrenholt, and S. {\AA}berg},
\bibinfo{journal}{Phys. Rev. C}
\textbf{\bibinfo{volume}{79}},
\bibinfo{pages}{064304}
(\bibinfo{year}{2009}).

\bibitem{Berry1981-87}
\bibinfo{author}{H.-J. Yuh, G.S. Ezra, P. Rehmus, and R.S. Berry},
\bibinfo{journal}{Phys. Rev. Lett.}
\textbf{\bibinfo{volume}{47}},
\bibinfo{pages}{497}
(\bibinfo{year}{1981});
\bibinfo{author}{G.S. Ezra and R.S. Berry},
\bibinfo{journal}{Phys. Rev. A}
\textbf{\bibinfo{volume}{28}},
\bibinfo{pages}{1974}
(\bibinfo{year}{1983});
\bibinfo{author}{J.L. Krause and R.S. Berry},
\bibinfo{journal}{\emph{ibid}}
\textbf{\bibinfo{volume}{31}},
\bibinfo{pages}{3502}
(\bibinfo{year}{1985});
\bibinfo{author}{J.E. Hunter III and R.S. Berry},
\bibinfo{journal}{\emph{ibid}}
\textbf{\bibinfo{volume}{36}},
\bibinfo{pages}{3042}
(\bibinfo{year}{1987}).

\bibitem{Matyus2011}
\bibinfo{author}{E. M\'atyus, J. Hutter, U. M\"uller-Herold, and M. Reiher},
\bibinfo{journal}{Phys. Rev. A}
\textbf{\bibinfo{volume}{83}},
\bibinfo{pages}{052512}
(\bibinfo{year}{2011});
\bibinfo{journal}{J. Chem. Phys.}
\textbf{\bibinfo{volume}{135}},
\bibinfo{pages}{204302}
(\bibinfo{year}{2011}).

\bibitem{Becerra2013}
\bibinfo{author}{M. Becerra, V. Posligua, and E.V. Lude\~na},
\bibinfo{journal}{Int. J. Quant. Chem.}
\textbf{\bibinfo{volume}{113}},
\bibinfo{pages}{1584}
(\bibinfo{year}{2013});
\bibinfo{author}{C.G. Rodr\'iguez, A.S. Urbina, F.J. Torres, D. Cazar, and E.V. Lude\~na}
\bibinfo{journal}{Comp. Theor. Chem.}
\textbf{\bibinfo{volume}{1018}},
\bibinfo{pages}{26}
(\bibinfo{year}{2013}).

\bibitem{Birman2013}
\bibinfo{author}{J.L. Birman, R.G. Nazmitdinov, and V.I. Yukalov},
\bibinfo{journal}{Phys. Rep.}
\textbf{\bibinfo{volume}{526}},
\bibinfo{pages}{1}
(\bibinfo{year}{2013}).

\bibitem{Berry1983}
\bibinfo{author}{R.S. Berry, G.S. Ezra, and G. Natanson, in}
\emph{\bibinfo{title}{New Horizons of Quantum Chemistry}},
edited by \bibinfo{author}{P.-O. L\"owdin and B. Pullman}
(\bibinfo{publisher}{D. Reidel Publishing Company, Dordrecht},
\bibinfo{year}{1983}) p 77.

\bibitem{Kellman1997}
\bibinfo{author}{M.E. Kellman},
\bibinfo{journal}{Int. J. Quant. Chem.}
\textbf{\bibinfo{volume}{65}},
\bibinfo{pages}{399}
(\bibinfo{year}{1997}).

\bibitem{Essen1977}
\bibinfo{author}{H. Ess\'en},
\bibinfo{journal}{Int. J. Quant. Chem.}
\textbf{\bibinfo{volume}{12}},
\bibinfo{pages}{721}
(\bibinfo{year}{1977}).

\bibitem{Villars1987}
\bibinfo{author}{F.M.H. Villars},
\bibinfo{journal}{Nucl. Phys. A}
\textbf{\bibinfo{volume}{473}},
\bibinfo{pages}{539}
(\bibinfo{year}{1987}).

\bibitem{Richard1992}
\bibinfo{author}{J.M. Richard},
\bibinfo{journal}{Phys. Rep.}
\textbf{\bibinfo{volume}{212}},
\bibinfo{pages}{1}
(\bibinfo{year}{1992}).

\bibitem{Hunter1975}
\bibinfo{author}{G. Hunter},
\bibinfo{journal}{Int. J. Quant. Chem.}
\textbf{\bibinfo{volume}{9}},
\bibinfo{pages}{237}
(\bibinfo{year}{1975}).

\bibitem{Hylleraas1929-64}
\bibinfo{author}{E.A. Hylleraas},
\bibinfo{journal}{Z. Phys.}
\textbf{\bibinfo{volume}{54}},
\bibinfo{pages}{347}
(\bibinfo{year}{1929});
\bibinfo{journal}{Adv. Quantum Chem.}
\textbf{\bibinfo{volume}{1}},
\bibinfo{pages}{1}
(\bibinfo{year}{1964}).

\bibitem{Gross2012-14}
\bibinfo{author}{A. Abedi, N.T. Maitra, and E.K.U. Gross},
\bibinfo{journal}{J. Chem. Phys.}
\textbf{\bibinfo{volume}{137}},
\bibinfo{pages}{22A530}
(\bibinfo{year}{2012});
\bibinfo{author}{N.I. Gidopoulos and E.K.U. Gross},
\bibinfo{journal}{Phil. Trans. R. Soc. A}
\textbf{\bibinfo{volume}{372}},
\bibinfo{pages}{20130059}
(\bibinfo{year}{2014}).

\bibitem{Arce2012}
\bibinfo{author}{J.C. Arce},
\bibinfo{journal}{Phys. Rev. A}
\textbf{\bibinfo{volume}{85}},
\bibinfo{pages}{042108}
(\bibinfo{year}{2012}).

\bibitem{Hunter1981}
\bibinfo{author}{G. Hunter},
\bibinfo{journal}{Int. J. Quant. Chem.}
\textbf{\bibinfo{volume}{19}},
\bibinfo{pages}{755}
(\bibinfo{year}{1981}).

\bibitem{Jecko2015}
\bibinfo{author}{T. Jecko, B.T. Sutcliffe, and R.G. Woolley},
\bibinfo{journal}{J. Phys. A: Math. Theor.}
\textbf{\bibinfo{volume}{48}},
\bibinfo{pages}{445201}
(\bibinfo{year}{2015}).

\bibitem{Cederbaum2013-2014}
\bibinfo{author}{L.S. Cederbaum},
\bibinfo{journal}{J. Chem. Phys.}
\textbf{\bibinfo{volume}{138}},
\bibinfo{pages}{224110}
(\bibinfo{year}{2013});
\textbf{\bibinfo{volume}{141}},
\bibinfo{pages}{029902}
(\bibinfo{year}{2014}).

\bibitem{Nakashima2007}
\bibinfo{author}{H. Nakashima and H. Nakatsuji},
\bibinfo{journal}{J. Chem. Phys.}
\textbf{\bibinfo{volume}{127}},
\bibinfo{pages}{224104}
(\bibinfo{year}{2007}).

\bibitem{Mathematica}
\bibinfo{author}{Wolfram Research, Inc.},
\emph{\bibinfo{title}{Mathematica, Version 10.3.1}}
\bibinfo{year}{(2015)}.

\bibitem{Bunker1998}
\bibinfo{author}{P. R. Bunker and P. Jensen},
\bibinfo{title}{\emph{Molecular Symmetry and Spectroscopy}, 2nd ed.}
(\bibinfo{publisher}{NRC Research Press, Ottawa},
\bibinfo{year}{1998}).

\bibitem{Koga2006}
\bibinfo{author}{T. Koga and H. Matsuyama},
\bibinfo{journal}{Theor. Chem. Acc.}
\textbf{\bibinfo{volume}{115}},
\bibinfo{pages}{59}
(\bibinfo{year}{2006});
\bibinfo{author}{A.W. King, L.C. Rhodes and H. Cox},
\bibinfo{journal}{Phys. Rev. A}
\textbf{\bibinfo{volume}{93}},
\bibinfo{pages}{022509}
(\bibinfo{year}{2016}).

\bibitem{Kais2000}
\bibinfo{author}{S. Kais and Q. Shi},
\bibinfo{journal}{Phys. Rev. A}
\textbf{\bibinfo{volume}{62}},
\bibinfo{pages}{060502-1}
(\bibinfo{year}{2000});
\bibinfo{author}{Q. Shi and S. Kais},
\bibinfo{journal}{Int. J. Quant. Chem.}
\textbf{\bibinfo{volume}{85}},
\bibinfo{pages}{307}
(\bibinfo{year}{2001}).


\bibitem{Boyd1982}
\bibinfo{author}{R.J. Boyd and M.C. Yee},
\bibinfo{journal}{J. Chem. Phys.}
\textbf{\bibinfo{volume}{77}},
\bibinfo{pages}{3578}
(\bibinfo{year}{1982}).

\bibitem{Hunter1986}
\bibinfo{author}{G. Hunter and C.C. Tai},
\bibinfo{journal}{Int. J. Quant. Chem. Symp.}
\textbf{\bibinfo{volume}{19}},
\bibinfo{pages}{173}
(\bibinfo{year}{1986}).

\bibitem{Feagin1986-88}
\bibinfo{author}{J.M. Feagin and J.S. Briggs},
\bibinfo{journal}{Phys. Rev. Lett.}
\textbf{\bibinfo{volume}{57}},
\bibinfo{pages}{984}
(\bibinfo{year}{1986});
\bibinfo{journal}{Phys. Rev. A}
\textbf{\bibinfo{volume}{37}},
\bibinfo{pages}{4599}
(\bibinfo{year}{1988}).

\bibitem{Hunter1967}
\bibinfo{author}{G. Hunter and H.O. Pritchard},
\bibinfo{journal}{J. Chem. Phys.}
\textbf{\bibinfo{volume}{46}},
\bibinfo{pages}{2146}
(\bibinfo{year}{1967});
\textbf{\bibinfo{volume}{46}},
\bibinfo{pages}{2153}
(\bibinfo{year}{1967})

\bibitem{Lin1995}
\bibinfo{author}{C.D. Lin},
\bibinfo{journal}{Phys. Rep.}
\textbf{\bibinfo{volume}{257}},
\bibinfo{pages}{1}
(\bibinfo{year}{1995}).

\end{thebibliography}

\end{document}